# The Relationship Between the Expansion Speed and Radial Speed of CMEs Confirmed Using Quadrature Observations of the 2011 February 15 CME

Nat Gopalswamy[1], Pertti Mäkelä[2], Seiji Yashiro[2], and Joseph M. Davila[1]

[1]NASA Goddard Space Flight Center, Greenbelt, Maryland, USA

[2]The Catholic University of America, Washington DC, USA

E-mail: nat.gopalswamy@nasa.gov

It is difficult to measure the true speed of Earth-directed CMEs from a coronagraph located along the Sun-Earth line because of the occulting disk. However, the expansion speed (the speed with which the CME appears to spread in the sky plane) can be measured by such a coronagraph. In order to convert the expansion speed to radial speed (which is important for space weather applications) one can use an empirical relationship between the two that assumes an average width for all CMEs. If we have the width information from quadrature observations, we can confirm the relationship between expansion and radial speeds derived by Gopalswamy et al. (2009a). The STEREO spacecraft were in qudrature with SOHO (STEREO-A ahead of Earth by 87° and STEREO-B 94° behind Earth) on 2011 February 15, when a fast Earth-directed CME occurred. The CME was observed as a halo by the Large-Angle and Spectrometric Coronagraph (LASCO) on board SOHO. The sky-plane speed was measured by SOHO/LASCO as the expansion speed, while the radial speed was measured by STEREO-A and STEREO-B. In addition, STEREO-A and STEREO-B images provided the width of the CME, which is unknown from Earth view. From the SOHO and STEREO measurements, we confirm the relationship between the expansion speed ($V_{exp}$) and radial speed ($V_{rad}$) derived previously from geometrical considerations (Gopalswamy et al. 2009a): $V_{rad} = ½ (1 + \cot w)V_{exp}$, where w is the half width of the CME. STEREO-B images of the CME, we found that CME had a full width of 76°, so $w = 38°$. This gives the relation as $V_{rad} = 1.14 V_{exp}$. From LASCO observations, we measured $V_{exp} = 897$ km/s, so we get the radial speed as 1023 km/s. Direct measurement of radial speed yields 945 km/s (STEREO-A) and 1058 km/s (STEREO-B). These numbers are different only by 7.6% and 3.4% (for STEREO-A and STEREO-B, respectively) from the computed value.

Keywords: coronal mass ejections, expansion speed, radial speed

## Introduction

Coronal mass ejections (CMEs) from the Sun influence the conditions in the magnetosphere, ionosphere, and the atmosphere via the impact of energetic plasma and particles associated with the CMEs. The energetic particles are accelerated by CME-driven shocks to very high energies throughout the interplanetary medium. Accelerated protons can get trapped in the radiation belts, produce excess ionization in the polar ionosphere, and even penetrate the atmosphere altering the atmospheric chemistry. When the CME–driven shock arrives at the magnetosphere, it can cause a sudden commencement (SC), often followed by an intense geomagnetic storm, which also has implications throughout the geospace and even on the ground. Thus, understanding the structure and kinematics of CMEs is important in predicting the CME travel time to Earth and the nature of geomagnetic storms they produce. Of particular interest is the estimate of the true speed of CMEs near the Sun based on measurements from coronagraphs located along the Sun-Earth line. CMEs directed toward Earth are not well observed by such coronagraphs because of a basic instrument requirement: an occulting disk that blocks the bright photosphere. Unfortunately, the occulting disk also blocks the CME nose portion that arrives at Earth. Therefore, what we measure in the coronagraph images is the flanks of the CME and/or shock. The CME portion detected in situ by spacecraft such as ACE and Wind, therefore, does not correspond to what is observed near the Sun by coronagraphs. After the advent of the Solar Terrestrial Relations Observatory (STEREO) mission, it became possible to observe Earth-directed CMEs without any obstruction, when STEREO spacecraft were in quadrature with the SOHO spacecraft. Under such configuration, CMEs can be observed in three views (STEREO Ahead, STEREO Behind, and SOHO) so we can infer the three-dimensional structure of the Earth-directed CMEs.

It is likely that we need to be content with CME observations from the Sun-Earth line until that time when observations by spacecraft stationed at Sun-Earth L5 or L4 become possible in the future (Gopalswamy e al. 2011a,b). In order to use the sky-plane speeds to determine the radial speeds, one can use a cone model to approximate the overall structure of the CMEs and hence obtain deprojected speeds (see e.g., Xie et al., 2004). Alternatively, one can establish an empirical relation between the expansion speed (the speed with which the lateral size of the CMEs increase) and the radial speed (the speed of the CME nose) (Dal Lago et al. 2003; Schwenn et al., 2005). Both of these models, however, require information on the width of the CMEs (Gopalswamy et al., 2009a), which is not available from observations made from the Sun-Earth line. Gopalswamy et al. (2010) described a method to estimate the CME width using the width – speed relationship of a large number of limb CMEs. For limb CMEs, the measured

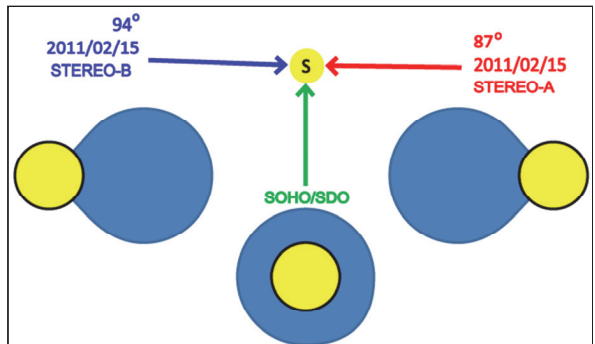

Fig.1. Schematic diagram showing the viewing directions from STEREO-A, Earth, and STEREO-B towards the Sun (S). The corresponding coronagraphic images are also shown (the disk represents the coronagraphic occulting disk). In Earth view (SOHO/SDO), the CME appears as a full halo. In STEREO-A (B) view, the CME appears above the east (west) limb of the Sun. The angles of STEREO A and B away from the Sun-Earth line are shown for the day 2011 February 11.

speed and width are close to the actual ones, but these CMEs do not arrive at Earth. This relationship can be used to obtain the approximate widths of Earth-directed CMEs. The empirical relation between the radial ($V_{rad}$) and expansion ($V_{exp}$) speeds derived by Dal Lago et al. (2003) and Schwenn et al. (2005) is

$V_{rad} = 0.88 V_{exp}$,  (1)

which does not involve the width dependence. Gopalswamy et al. (2009a) derived a $V_{rad}$ -$V_{exp}$ relation from the geometrical properties of CMEs assumed to have the structure of an ice cream cone:

$V_{rad} = f(w) V_{exp}$,  (2)

where the function $f$ depends on the CME half width $w$ and on the exact definition of the CME cone. They found that the full ice-cream cone, for which

$f(w) = ½ (1 + \cot w)$,  (3)

represents the data best. This was also confirmed by Michalek et al. (2009) using a large number of CMEs originating close to the limb. In this paper, we confirm the Gopalswamy et al. (2009a) relationship for the case of an Earth-directed CME on 2011 February 15 observed as a full halo by SOHO/LASCO and simultaneously by STEREO-A and B. The CME was observed as a limb event in both STEREO A and B and hence $w$ and $V_{rad}$ can be directly measured. The SOHO/LASCO observations provide $V_{exp}$, thus providing all the information needed to confirm the $V_{rad}$ -$V_{exp}$ relation shown in equation (2).

**Observations**

The 2011 February 15 CME is associated with the first X-class flare of solar cycle 24 and one of several CMEs from the active region with NOAA Number 1158 (Schrijver et al., 2011). The X2.2 flare occurred from S20W11, starting, peaking, and ending at 01:46:50, 01:54:08 and 03:37:39 UT, respectively. The CME appeared first in the SOHO/LASCO field of view (FOV) at 02:24 UT above the southwest limb and soon developed into a full halo CME by 02:48 UT, consistent with the location of the source region close to the disk center. Figure 1 shows the quadrature configuration of the STEREO and SOHO spacecraft on 2011 February 15: STEREO A was ~87° west of the Sun-Earth line, while STERE-B was ~94° to the east. The halo CME in Earth view was observed as a limb CME by the SECCHI coronagraphs (Howard et al., 2008) in both STEREO-A and B views.

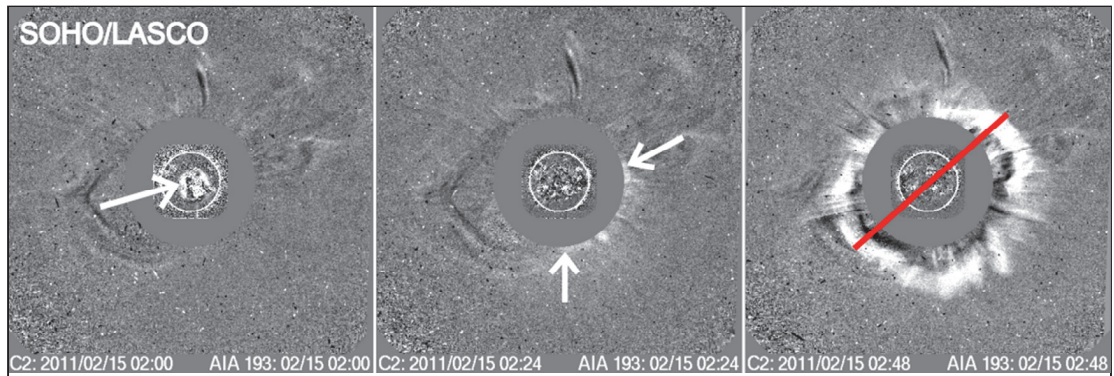

Fig.2. Three SOHO/LASCO snapshot images showing the evolution of the 2011 February 15 CME. The images are superposed with SDO/AIA 193 Å images to show the disk location of the eruption (pointed by arrow in the 02:00 image). The disturbance appears above the C2 occulting disk at 20:24 UT (between the two arrows). At 02:48 UT, the CME surrounds the occulting disk as a full halo. The solid line connects the leading edges of the CME at position angles 120° and 300°, along which the expansion speed (the rate of increase of the length of the solid line) was measured.

Figure 2 shows the appearance and early evolution of the CME until it became a full halo in the SOHO/LASCO/C2 images. These are running difference images showing the CME evolution. The images also show 193 Å EUV difference images from the Atmospheric Imaging Assembly (AIA) on board the Solar Dynamics Observatory (SDO) superposed. A large-scale eruption is already under way at 02:00 UT and appears above the C2 occulting disk 24 min later (between the two arrows in the middle panel). Finally the CME appears as a full halo by 20:48 UT.

In the STEREO-A and B views, the CME appeared as a limb event as shown in Fig. 3. In both views, the limb CME covers a position angle extent of ~76°. The CME looks very similar in the two views and has the bright flux rope-CME surrounded by a diffuse structure, which is the sheath of the CME-driven shock (see Gopalswamy and Yashiro, 2011 and references therein). The CME-height-time measurements were made at the nose of the shock shown circled in the images.

Figure 4 shows the height-time plots obtained using measurements from SOHO and STEREO images. From the LASCO images, the CME heights were measured at two diametrically opposite position angles and then combined to get the expansion speed. In fact we measured the CME speed at 10 position angles around the Sun. The speeds varied from 335 km/s along PA = 50° to 739 km/s along PA = 225°. Note that the CME erupted from the southwest quadrant, so there is slight asymmetry in the halo appearance and hence the difference in speed in these position angles. However, the speed with which the diameter of the CME increased is roughly the same, similar to the one shown in Fig. 4. The reason we chose PA = 120° and 300° is that the CME was quite symmetric in this direction. The maximum speed was found near the equator in both views despite the S20 latitude of the source region (945 km/s in STEREO-A and 1058 km/s in STEREO B). This seems to be due to the deflection of the CME by a large coronal hole at the south pole. There was also a small corridor in the coronal hole, which seems to be a quiet-Sun region and the shock might have propagated through it causing the hook-like sharp feature in the southern end of the CME. These features will be reported elsewhere.

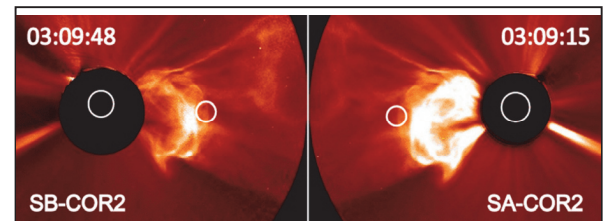

Fig. 3. STERE-B (left) and STEREO-A (right) images of the 2011 February 15 CME by the COR2 coronagraph of the SECCHI instrument package. The circles mark the nose part of the CME at which the height-time measurements were made. The white circles on the occulting disk mark the optical size of the Sun. The full width of the CME is ~76° in both view.

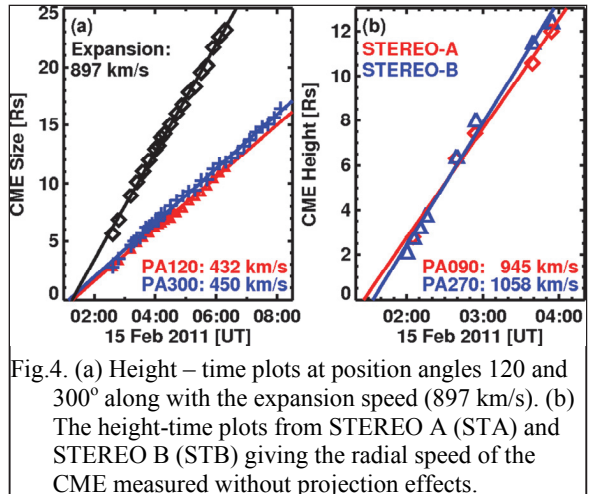

Fig.4. (a) Height – time plots at position angles 120 and 300° along with the expansion speed (897 km/s). (b) The height-time plots from STEREO A (STA) and STEREO B (STB) giving the radial speed of the CME measured without projection effects.

**Results**

The height-time measurements provide direct information on the expansion speed (897 km/s) and radial speed (945 km/s and 1058 km/s in STEREO A and B, respectively). We can now check these values against the theoretical values obtained from geometrical considerations (equation (2)). From Fig. 3, we obtain the full width of the CME in the sky plane as 76°. Since the CME was observed as a limb event by STEREO A and B, the projection effects are minimal, so the measured width is close to the true width. From this we obtain the half width of the CME cone as $w=38°$, as illustrated in the schematic in Fig. 5a. When we substitute $w = 38°$ in equation (3), we get $f(w) = 1.14$ as shown in Fig. 5b. Once we get $f(w)$, we can get the radial speed from the expansion speed as $V_{rad} = 1.14\ V_{exp}$ from equation (2). If we substitute the expansion speed from LASCO observations ($V_{exp} = 897$ km/s), we get $V_{rad} = 1023$ km/s. We have independent measurements of the radial speed from the STEREO observations, which are 945 km/s (STEREO-A) and 1058 km/s (STEREO-B). The measured radial speeds are very close to the value given by the full ice-cream cone model, deviating only 7.6% for STEREO A and 3.4% for STEREO B. These deviations are well within the typical errors in height-time measurements (~10%). Thus we conclude that the relationship between radial speed and expansion speed derived in Gopalswamy et al. (2009a) using the full ice-cream cone model is directly validated using the multiple-view observations of the 2011 February 15 CME.

When we use the flat cone model (see Fig. 5b), we get $f = 0.64$. In this case the radial speed will be smaller than the expansion speed: 574 km/s (see equation (2)). This definitely an underestimate because it is even smaller than the sky-plane speed (669 km/s) measured in the LASCO field of view. The deviation from the measured value (e.g. 1058 km/s in STEREO-B) is ~46%. For the partial ice-cream cone model in Fig.5b, we get $f = 0.81$, so the radial speed is smaller than the expansion speed: 728 km/s, which is ~29% smaller than the measured value. Finally, the constant value of $f=0.88$ (see equation (1)) obtained by Schwenn et al. (2005) also gives a radial speed (789 km/s) smaller than the expansion speed and deviates from the measured value (1058 km/s) by ~25%. These results further confirm the validity of the full ice-cream cone model from direct measurements, consistent with the statistical result obtained by Michalek et al. (2009). The 25% deviation of the empirical relationship (1) also points to the importance of CME width in the speed relationship.

**Discussion and Conclusions**

The results presented in this paper point to the importance of observing the Earth-directed CMEs from a vantage point different from the Sun-Earth line. Without the STEREO data, the 2011 February CME would be classified as a low-speed (669 km/s) halo CME. The STEREO view provided an excellent opportunity to obtain the true speed of the CME heading toward Earth. We also saw that the speeds obtained from STEREO A and STEREO B were nearly identical because of their locations ~90° away from the Sun-Earth line. Of course the information is redundant – either one of the two STEREO spacecraft would have been sufficient to get the true speed toward Earth. This suggests the importance of having an observatory at Sun-Earth L5 or L4 in observing Earth-directed CMEs. Since the Sun rotates from east to west, it may be more advantageous to have an observatory at L5 because one can observe active regions before they rotate into Earth view.

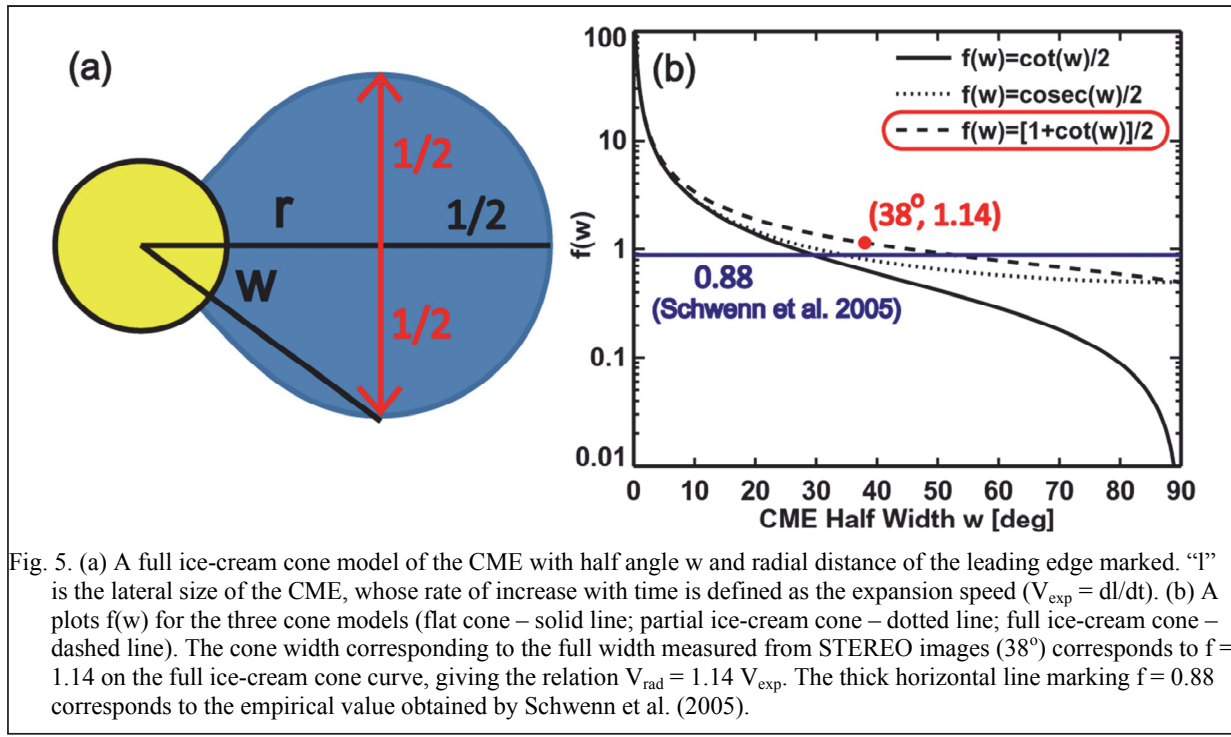

Fig. 5. (a) A full ice-cream cone model of the CME with half angle w and radial distance of the leading edge marked. "l" is the lateral size of the CME, whose rate of increase with time is defined as the expansion speed ($V_{exp} = dl/dt$). (b) A plots f(w) for the three cone models (flat cone – solid line; partial ice-cream cone – dotted line; full ice-cream cone – dashed line). The cone width corresponding to the full width measured from STEREO images (38°) corresponds to f = 1.14 on the full ice-cream cone curve, giving the relation $V_{rad} = 1.14\, V_{exp}$. The thick horizontal line marking f = 0.88 corresponds to the empirical value obtained by Schwenn et al. (2005).

We have used just the STEREO/COR2 observations in this analysis. The main reason is that the COR2 field of view is similar to that of LASCO (C2 and C3). More detailed information is available from the EUVI instruments and the inner coronagraph COR1 to observe the CME much closer to the surface. The COR1 speeds measured were not too different, but slightly higher: 1321 km/s (STEREO-A) and 1275 km/s (STEREO-B). The maximum speeds were also at slightly different position angles in the COR1 field of view: 100° (STEREO-A) and 260° (STEREO-B). Note that these angles correspond to directions slightly south of the ecliptic plane, indicating that the CME was moving toward the equator because of the deflection by the coronal hole (Gopalswamy et al., 2009b,c) at the south pole. The COR2 speeds peaked near the ecliptic plane, amounting to a deflection of ~20°. Deflections of this magnitude were common during the rise phase of cycle 23 (Gopalswamy et al., 2003).

Note that we did not distinguish between shock speed and the CME speed in this work. This may not be significant near the Sun, but the distinction is important far away from the sun, where the shock standoff distance gets very large. True Earthward speeds obtained in the coronagraphic field of view are important to get days of advance warning on CMEs heading toward Earth. Such an advanced warning is very important for space weather forecasting. However, it must be pointed out that the usage of heliospheric imagers provides useful information on the continued propagation of CMEs that help characterize the interaction with the background solar wind (Möstl et al., 2011).

We also confirm that the Earthward speed of CMEs could in principle be obtained from the expansion speed of halo CMEs measured in the images obtained by a coronagraph located along the Sun-Earth line, provided the CME width can be estimated by an independent means. Gopalswamy et al. (2010) made use of the relation between speed and width of limb CMEs to obtain a rough estimate of CME half width ($w$) depending on the sky-plane speed range: 66° ($V_{sky}$ > 900 km/s), 45° (500 km/s < $V_{sky}$ ≤ 900 km/s), and 32° ($V_{sky}$ < 500 km/s). The sky-plane speed in the direction of the fastest moving section was obtained as 669 km/s (see the CME catalog, http://cdaw.gsfc.nasa.gov/CME_list). The corresponding half width is 45°, which is slightly higher than the true width (38°) measured by the STEREO coronagraphs. When we use $w$ = 45° in equation (3), we get $f(w)$ = 1, slightly smaller than the value obtained from direct measurements (1.14). Accordingly, the radial speed will be the same as the expansion speed, which is about 12% smaller than the value obtained from direct measurements. The approximate scheme presented in Gopalswamy et al. (2010) seems to be reasonable at least for this CME.

To conclude, the expansion and radial speeds of the 2011 February 15 CME obtained using SOHO and STEREO quadrature observations confirm the theoretical relationship of Gopalswamy et al (2009a) based on geometrical considerations of a full ice-cream cone model of the CME. Although the CME width can be crudely estimated from the empirical relationship between CME width and speed for limb CMEs, we suggest that measuring the Earthward speed of CMEs from a view away from the Sun-Earth line is the best option that is highly relevant to space weather forecasting. Finally, we appreciate what we presented is a case study. A similar analysis will be performed for a large number of events observed in quadrature and the statistical results will be reported elsewhere.

## Acknowledgments

Work supported by NASA's LWS TR&T program. The SECCHI instrument was constructed by a consortium of international institutions: the Naval Research Laboratory (USA), the Lockheed Martin Solar and Astrophysical Laboratory (USA), the NASA Goddard Space Flight Center (USA), the Max Planck Institut für Sonnensystemforschung (Germany), the Centre Spatial de Liège (Belgium), the University of Birmingham (UK), the Rutherford Appleton Laboratory (UK), the Institut d'Optique (France), and the Institute d'Astrophysique Spatiale (France). SOHO is a project of international cooperation between ESA and NASA.